\documentclass[12pt]{article}

\usepackage{amssymb}
\usepackage{amsmath}
\usepackage{latexsym}
\usepackage{yfonts}

\catcode`@=11
\renewcommand{\section}{\@startsection{section}{1}{0pt}{\medskipamount}
{\medskipamount}{\large\bf}} \numberwithin{equation}{section}
\catcode`@=12

\def\beq{\begin{eqnarray}}    
\def\eeq{\end{eqnarray}}      

\def\ln{\,\mbox{ln}\,}                  

\def\pa{\partial}                       

\def\={\ =\ }

\def\de{\delta}

\def\Ga{\Gamma}

\begin{document}

\begin{center}

{\Large\bf Basic properties of an alternative flow equation in gravity theories}

\vspace{18mm}

{\large
Peter M. Lavrov$^{(a, b)}\footnote{E-mail:
lavrov@tspu.edu.ru}$\; }

\vspace{8mm}

\noindent  ${{}^{(a)}} ${\em
Tomsk State Pedagogical University,\\
Kievskaya St.\ 60, 634061 Tomsk, Russia}

\noindent  ${{}^{(b)}} ${\em
National Research Tomsk State  University,\\
Lenin Av.\ 36, 634050 Tomsk, Russia}

\vspace{20mm}

\begin{abstract}
\noindent
Basic properties  of alternative flow equation in Quantum Gravity
are studied. It is shown that the alternative flow equation
for effective two-particle irreducible effective
action is  gauge independent  and does not depend on IR parameter
$k$ on shell.

\end{abstract}

\end{center}

\vfill

\\

\newpage

\section{Introduction}
\noindent
Being popular among the functional renormalization group (FRG) community
the standard formulation
of the FRG approach
\cite{Wet1,Wet2} (among recent reviews of the method see, for example, \cite{DCEMPTW})
meets with the very serious  gauge dependence problem
of the effective average action as  within  the  perturbation theory \cite{LSh}
as well as on the level
 when it is found as a  non-perturbative
solution to the flow equation making  impossible
physical interpretations of any results obtained in gauge theories with
the help of effective average action \cite{Lav-2020,Lav-yad}.

Recently it has been proposed \cite{AMNS-1} and
studied \cite{Lav-alt}  alternative
methods in comparison with \cite{LSh} when the regulators being
essential tools of the FRG are considered as sources to composite operators.
It differs with the standard introduction of external sources
to composite operators
\cite{CJT} (generalizations to the case of gauge theories
have been proposed in \cite{LO-1989,Lav-tmph,LOR-jmp})
when they are usual functions of space-time coordinates
instead of the regulators which are some differential operators
in general \cite{Reuter}. In contrast with the standard FRG \cite{Wet1,Wet2}
the alternative methods \cite{AMNS-1,Lav-alt} lead to the 2PI effective average
actions which have good properties within the perturbation theory. Unfortunately
they still unacceptable on the non-perturbative level because solutions to corresponding
flow equations depend on gauge at any value of the IR parameter $k$ including the fixed
points \cite{Lav-alt}.

In the present  paper we are going to extend the reformulation of the FRG proposed in
\cite{LSh} for Yang-Mills fields to the case of gravity theories  and
to study basic properties of corresponding flow equation.
The paper is organized as follows.
In Sec. 2 the properties of effective action on an arbitrary background metric
for gravity theories in
the standard quantization scheme \cite{FP} are presented.
In Sec. 3 the gauge dependence problem
of 2PI irreducible effective average action for Quantum Gravity within  the perturbation
theory is discussed.
In Sec. 4  the derivation of alternative flow equation for the
effective action with composite operators and the study its $k$-dependence are given.
In Sec. 5 the gauge dependence of the alternative flow equations
is investigated.
Finally, in Sec. 6 the results obtained in the paper are discussed.

The DeWitt's condensed notations \cite{DeWitt} are used.
 The functional derivatives with respect
to fields and sources
are considered as right and left correspondingly.   The left
functional derivatives with respect to
fields  are marked by special symbol $"\rightarrow"$.
Arguments of any functional are enclosed in square brackets
$[\;]$,
and arguments of any function are enclosed in parentheses, $(\;)$.
The symbol $F_{,i}[\phi,...]$ means the
right derivative of $F[\phi,...]$ with respect to field $\phi^i$.

\section{Effective action for gravity theories}
\noindent
We start with an arbitrary initial action $S_0[g]$ of the metric tensor
$g=\{g_{\mu\nu}\}$.\footnote{Standard examples are Einstein gravity,
$S_0(g)=\kappa^{-2}\int dx \sqrt{-{\rm det}g}\;\!R$,
and $R^2$ gravity,
$S_0(g)=\int dx \sqrt{-{\rm det}g}\;(\lambda_1 R^2+
\lambda_2R^{\mu\nu}R_{\mu\nu}+\kappa^{-2}R)$.}
We suppose the invariance of $S_0[g]$,
\beq
\label{A1}
\delta_{\xi}S_0[g]=0,
\eeq
under general coordinate transformations which  infinitesimally  take the
form of gauge transformations of $g_{\mu\nu}$
\beq
\label{A2}
\delta_{\xi} g_{\mu\nu}=-\pa_{\sigma}g_{\mu\nu}\xi^{\sigma}-
g_{\mu\sigma}\pa_{\nu}\xi^{\sigma}-g_{\sigma\nu}\pa_{\mu}\xi^{\sigma}=
R_{\mu\nu\sigma}(g)\xi^{\sigma},
\eeq
where $\xi^{\sigma}$ are arbitrary functions of space-time coordinates and
$R_{\mu\nu\sigma}(g)$ are the gauge generators satisfying the closed
and irreducible gauge algebra (for detailed description see \cite{BLRNSh}),
\beq
\label{A3}
[\delta_{\xi_1},\delta_{\xi_2}]g_{\mu\nu}=\delta_{\xi_3}g_{\mu\nu},\quad
\xi^{\sigma}_3=F^{\sigma}_{\alpha\beta}\xi^{\beta}_2\xi^{\alpha}_1=
\xi^{\sigma}_1\pa_{\alpha}\xi^{\alpha}_2-
\xi^{\sigma}_2\pa_{\alpha}\xi^{\alpha}_1.
\eeq
In (\ref{A3}) $F^{\sigma}_{\alpha\beta}$ are structure coefficients of gauge
algebra which do not depend on fields $g_{\mu\nu}$
and have the universal form for any gravity
theory.

On quantum level one operates with the action
\beq
\label{A4}
S[\phi,{\bar g}]=S_0[h+{\bar g}]+S_{gh}[\phi,{\bar g}]+S_{gf}[\phi,{\bar g}],
\eeq
appearing in the Faddeev-Popov method \cite{FP}.
Here the decomposition of $g$, $g=h+{\bar g}$,
on a background metric ${\bar g}=\{{\bar g}_{\mu\nu}\}$ and
quantum fluctuation $h=\{h_{\mu\nu}\}$ is used.
In (\ref{A4}) $\phi^i=(h_{\alpha\beta},B^{\alpha},C^{\alpha},{\bar
C}^{\alpha})$ is the set of quantum fields,  $C^{\alpha},{\bar
C}^{\alpha}$ are the ghost and antighost fields, $B^{\alpha}$ are
auxiliary Nakanishi-Lautrup fields for introducing the gauge fixing functions,
$\chi_{\alpha}({\bar g},h)$. A standard choice of $\chi_{\alpha}({\bar g}, h)$
corresponding to the background field
gauge condition in linear gauges  \cite{Barv} reads
\beq
\label{A5}
\chi_{\alpha}({\bar g}, h)={\cal F}^{\mu\nu}_{\alpha}({\bar g})h_{\mu\nu}, \quad
{\cal F}^{\mu\nu}_{\alpha}({\bar g})=
-{\bar g}^{\mu\sigma}\big(a\delta^{\nu}_{\alpha}
{\bar \nabla}_{\sigma}+b\delta^{\nu}_{\sigma}{\bar \nabla}_{\alpha}\big ),
\eeq
where ${\bar \nabla}_{\sigma}$ is the covariant derivative constructed with the help of
${\bar g}_{\mu\nu}$ and $a,b$ are constants. The de Donder gauge condition
corresponds to the case when $a=1,b=1/2$.
$S_{gh}[\phi,{\bar g}]$ is the ghost action,
\beq
\label{A6}
S_{gh}[\phi,{\bar g}]=\int dx\sqrt{-{\rm det}{\bar g}}\;
{\bar C}^{\alpha}G_{\alpha}^{\beta\gamma}({\bar g},h)
R_{\beta\gamma\sigma}({\bar g}+h)C^{\sigma},
\eeq
with  the notation
\beq
\label{A7}
G_{\alpha}^{\beta\gamma}({\bar g},h)=
\frac{\delta\chi_{\alpha}({\bar g},h)}{\delta h_{\beta\gamma}}\;,
\eeq
and $S_{gf}[\phi,{\bar g}]$ is the gauge fixing action
\beq
\label{A8}
S_{gf}[\phi,{\bar g}]=\int dx \sqrt{-{\rm det}{\bar g}}\;
B^{\alpha}\chi_{\alpha}({\bar g},h).
\eeq

For any admissible choice of gauge fixing functions
$\chi_{\alpha}({\bar g},h)$ the action
(\ref{A4}) is invariant under global supersymmetry (BRST symmetry)
\cite{BRS1,T}, \!\!\!
 \footnote{The gravitational BRST transformations were introduced in
 \cite{DR-M,Stelle,TvN}. For more compact presentation of the BRST transformations we
 use the notation $\delta_B$ for $\delta_{BRST}$.}
\beq
\label{A9}
\delta_B h_{\mu\nu}=R_{\mu\nu\alpha}({\bar g}+h)C^{\alpha}\Lambda,\quad
\delta_B B^{\alpha}=0,\quad
\delta_B C^{\alpha}=-C^{\sigma}\pa_{\sigma} C^{\alpha}\Lambda,\quad
\delta_B {\bar C}^{\alpha}=B^{\alpha}\Lambda,
\eeq
where $\Lambda$ is a constant Grassmann parameter.

The generating functional of Green functions, $Z=Z[J,{\bar g}]$, is constructed in the form of
functional integral
\beq
\label{A10}
Z[J,{\bar g}]=\int D\phi
\exp\Big\{\frac{i}{\hbar}\big(S[\phi,{\bar g}]+J\phi\big)\Big\}=
\exp\Big\{\frac{i}{\hbar}W[J,{\bar g}]\Big\},
\eeq
where $W=W[J,{\bar g}]$ is the generating functional of connected Green functions and
$J=\{J_i\}$ is the set of external sources to fields $\phi=\{\phi^i\}$. The
generating functional of vertex functions (effective action),
$\Gamma=\Gamma[\Phi, {\bar g}]$, is defined through the Legendre transform of $W$,
\beq
\label{A11}
\Gamma[\Phi, {\bar g}]=W[J,{\bar g}]-J\Phi,\quad
\Phi^i=\frac{\delta W}{\delta J_i}, \quad \frac{\delta\Gamma}{\delta\Phi^i}=-J_i
\eeq
and can be found as a solution to the following functional integro-differential equation,
\beq
\label{A12}
\exp\Big\{\frac{i}{\hbar}\Gamma[\Phi,{\bar g}]\Big\}=
\int D\phi
\exp\Big\{\frac{i}{\hbar}\Big(S[\Phi+\phi,{\bar g}]-
\frac{\delta\Gamma[\Phi, {\bar g}]}{\delta\Phi}\phi\Big)\Big\}.
\eeq
The standard approach (perturbation theory) in quantum field theory to find
a solution to the Eq. (\ref{A12}) is based on using
loop expansions
\beq
\Gamma[\Phi,{\bar g}]=S[\Phi,{\bar g}]+\hbar \Gamma_1[\Phi,{\bar g}]+...
\eeq
where
\beq
\Gamma_1[\Phi,{\bar g}]=-i{\rm Tr}\ln (S^{(2)}[\Phi,{\bar g}]), \quad
S^{(2)}[\Phi,{\bar g}]=\frac{\delta^2 S[\Phi,{\bar g}]}{\delta\Phi\;\delta\Phi},
\eeq
is one-loop approximation and ellipses mean higher order
loop contributions.

The generating functionals $Z,W,\Gamma$ depend on gauges but due to the BRST symmetry
the gauge dependence has a very special form and for variation $\delta\Gamma$
under an infinitesimal change of gauge fixing functions,
$\chi\rightarrow\chi+\delta\chi$,  obeys the property
\beq
\label{A13}
\delta_{\chi}\Gamma\Big|_{\frac{\delta\Gamma}{\delta\Phi}=0}=0,
\eeq
i.e. it does not depend on gauges when it is considered on extremals.
For the first time the gauge dependence of effective action
for gravity theories in the form
(\ref{A13}) has been described in \cite{LR} (for more early descriptions of gauge dependence
of effective action in gauge theories see papers \cite{J,Niel,LT-1981a,LT-1981b,VLT}).
This fact allows to state the gauge independence of $S$-matrix
thanks to the equivalence theorem \cite{KT}.
Among other important properties being very useful in practical calculations
within the background field method \cite{DeW,AFS,Abbott} the functionals
\beq
\label{A14}
Z[{\bar g}]=Z[J=0,{\bar g}],\quad
W[{\bar g}]=W[J=0,{\bar g}],\quad
\Gamma[{\bar g}]=\Gamma[\Phi=0, {\bar g}]
\eeq
are covariant functionals with respect to ${\bar g}$,
\beq
\label{A15}
\delta_{\xi}Z[{\bar g}]=0,\quad
\delta_{\xi}W[{\bar g}]=0,\quad
\delta_{\xi}\Gamma[{\bar g}]=0,
\eeq
as well as they do not depend on gauges
\beq
\label{A16}
\delta_{\chi}Z[{\bar g}]=0,\quad
\delta_{\chi}W[{\bar g}]=0,\quad
\delta_{\chi}\Gamma[{\bar g}]=0,
\eeq
as the direct consequences of the BRST invariance of action
$S[\phi,{\bar g}]$ \cite{LavSh-QG}.

\section{Gauge dependence of modified effective average action}
\noindent
The effective average action  of the FRG \cite{Wet1,Wet2}
is ill-defined  perturbatively in the case of Yang-Mills
theories \cite{LSh,Lav-yad} and gravity theories \cite{BLRNSh} because of
the gauge dependence
of effective average action  on-shell. To improve
the situation for Quantum Gravity we apply
the background field method \cite{DeW,AFS,Abbott}
\footnote{For recent development of the background field
method see \cite{GLSh}.}
and the formulation of effective action with
composite operators \cite{CJT,LO-1989,Lav-tmph} for construction of modified
effective average action
in the form used for the first time  in the
case of Yang-Mills theories in \cite{LSh}.

With the help of addition of a scale-dependent  regulator action, $S_k$,
being quadratic in the quantum fields, the FRG modifies
behavior of propagators of quantum fields in
IR and UV regions \cite{Wet1,Wet2}.
In the case of Quantum Gravity the scale-dependent
regulator action takes the form
\cite{Reuter}
\beq
\nonumber
S_{k}(\phi,{\bar g})&=&\int \!dx \sqrt{-{\rm det}{\bar g}}
\;\Big[\frac{1}{2}h_{\mu\nu}R^{(1)\mu\nu\;\!\alpha\beta}_{k}
({\bar g})h_{\alpha\beta}+
{\bar C}^{\alpha}R^{(2)}_{k \;\alpha\beta}({\bar g})C^{\beta}\Big]\equiv\\
\label{B1}
&\equiv& \int \!dx \sqrt{-{\rm det}{\bar g}}\big({\cal L}^{(1)}_k(h,{\bar g})+
{\cal L}^{(2)}_k(C,{\bar C},{\bar g})\Big),
\eeq
where $R^{(1)\mu\nu\;\!\alpha\beta}_{k}({\bar g})$,
$R^{(2)}_{k \;\alpha\beta}({\bar g})$
are regulators  with properties
\beq
\label{B2}
R^{(1)\mu\nu\;\alpha\beta}_{k}({\bar g})=
R^{(1)\;\!\alpha\beta\;\!\mu\nu}_{k}({\bar g}),\quad
\lim_{k\rightarrow 0}R^{(1)\mu\nu\;\alpha\beta}_{k}({\bar g})=0,\quad
\lim_{k\rightarrow 0}R^{(2)}_{k\; \alpha\beta}({\bar g})=0.
\eeq
On quantum level the FRG operates with the full action
\beq
\label{B3}
S_{Wk}[\phi,{\bar g}]=S[\phi,{\bar g}]+S_{k}[\phi,{\bar g}],
\eeq
where $S[\phi,{\bar g}]$ is defined in (\ref{A4}) - (\ref{A8}).
The action $S_{Wk}[\phi,{\bar g}]$ (\ref{B3}) is not invariant
under the BRST transformations (\ref{A9})
that leads to the gauge dependence
problem within the FRG  for Quantum Gravity \cite{BLRNSh}.

As it was already mentioned above
to improve the situation we propose
following to \cite{LSh} the generating functional of Green functions
$Z_k=Z_k[J,{\bar g},\Sigma]$
in the form
\beq
\label{B4}
Z_k=\int D\phi
\exp\Big\{\frac{i}{\hbar}\big(S[\phi,{\bar g}]+J\phi+
\Sigma{\cal L}_k(\phi,{\bar g})\big)\Big\},
\eeq
where $\Sigma=(\Sigma_1,\Sigma_2)$ are external sources to composite fields
${\cal L}_k(\phi,{\bar g})=({\cal L}^{(1)}_k(h,{\bar g}),
{\cal L}^{(2)}_k(C,{\bar C},{\bar g}))$ and $J=\{J_i\}$ are external
sources to fields
$\phi=\{\phi^i\}$. Let us note that from
the definition of generating functional $Z_k$ (\ref{B4})
and properties of  the regulators (\ref{B2})
it immediately follows coincidence of  the functional  $Z_k$
 with the standard generating functional of Green functions in the limit $k\rightarrow 0$.
The same statement is valid for the corresponding S-matrices.

Now we are going to study  dependence of the functional
$Z_k[J,{\bar g},\Sigma]$ (\ref{B4})
on gauges. To simplify the corresponding presentation it is useful to rewrite
the action $S[\phi,{\bar g}]$
in the form
\beq
\label{B5}
S[\phi,{\bar g}]=S_0[h+{\bar g}]+\Psi[\phi,{\bar g}]
{\hat R}[\phi,{\bar g}]
\eeq
with the help of gauge fixing functional $\Psi[\phi,{\bar g}]$
\beq
\label{B6}
\Psi[\phi,{\bar g}]=\int dx \sqrt{-{\rm det}{\bar g}}
\;{\bar C}^{\alpha}\chi_{\alpha}({\bar g},h),
\eeq
containing all information about gauge fixing,
and with the generator of BRST transformations (\ref{A9})
\beq
\label{B7}
{\hat R}[\phi,{\bar g}]=\int dx\;
\frac{\overleftarrow{\delta}}{\delta\phi^i}R^i(\phi, {\bar g}),\quad
R^i(\phi, {\bar g})=\big(R_{\mu\nu\sigma}({\bar g}+h)C^{\sigma},\; 0\;,
 -C^{\sigma}\pa_{\sigma} C^{\alpha}, B^{\alpha}\big).
\eeq
Consider an infinitesimal variation of gauge fixing functions,
$\chi_{\alpha}({\bar g},h)\;\rightarrow\;\chi_{\alpha}({\bar g},h)+
\delta\chi_{\alpha}({\bar g},h)$ which causes the variation of gauge fixing functional,
$\Psi[\phi,{\bar g}]\;\rightarrow \Psi[\phi,{\bar g}]+\delta\Psi[\phi,{\bar g}]$.
Let us temporally introduce the notations $S_{\Psi}[\phi,{\bar g}]=S[\phi,{\bar g}]$
and $Z_{k\Psi}=Z_k$
to stress essential dependence of $S[\phi,{\bar g}]$ and $Z_k$ on gauge fixing procedure.
In the functional integral
\beq
\label{B8}
Z_{k\Psi+\delta\Psi}=\int D\phi
\exp\Big\{\frac{i}{\hbar}\big(S_{\Psi}[\phi,{\bar g}]+
\delta\Psi[\phi,{\bar g}]{\hat R}[\phi,{\bar g}]+J\phi+
\Sigma{\cal L}_k(\phi,{\bar g})\big)\Big\},
\eeq
we make use the change of integration variables
in the form of BRST transformations
but trading the constant parameter $\Lambda$ by functional
\beq
\label{B9}
\Lambda[\phi,{\bar g}]=\frac{i}{\hbar}\delta\Psi[\phi,{\bar g}].
\eeq
Taking into account the corresponding Jacobian
\beq
\label{B10}
J[\phi,{\bar g}]=
\exp\Big\{-\frac{i}{\hbar}\delta\Psi[\phi,{\bar g}]{\hat R}[\phi,{\bar g}]\Big\},
\eeq
omitting the subscript $\Psi$ we obtain the following equation
\beq
\label{B11}
\delta Z_k
=\frac{i}{\hbar}\big(J_i+
\Sigma{\cal L}_{k,i}(\widehat{q},{\bar g})
\big)R^i(\widehat{q},{\bar g})
\delta\Psi[\widehat{q},{\bar g}]\;Z_k
\eeq
describing the gauge dependence of the functional $Z_k=Z_k[J,{\bar g},\Sigma]$.
In (\ref{B11}) the notations
\beq
\label{B12}
{\cal L}_{k,i}(\phi,{\bar g})=
\frac{\pa{\cal L}_{k}(\phi,{\bar g})}{\pa \phi^i},\quad
\widehat{q}^{\;i}=-i\hbar\frac{\delta}{\delta J_i}
\eeq
are used.
From (\ref{B11})  it follows the important statement that the gauge dependence
of $Z_k[J,{\bar g},\Sigma]$ disappears when  external sources are switched off,
$J_i=\Sigma_1=\Sigma_2=0$.

In terms of the generating functional of connected Green functions,
$W_k=W_k[J,{\bar g},\Sigma]=-i\hbar\ln Z_k[J,{\bar g},\Sigma]$,
the relation (\ref{B11}) takes the form
\beq
\label{B13}
\delta W_k=\big(J_i+
\Sigma{\cal L}_{k,i}(\widehat{Q}_k,{\bar g})
\big) R^i(\widehat{Q}_k,{\bar g})
\delta\Psi[\widehat{Q}_k,{\bar g}]\;\cdot  1,
\eeq
where
\beq
\label{B14}
\widehat{Q}^{\;i}_k=\widehat{q}^{\;i}+\frac{\delta W_k}{\delta J_i}.
\eeq

The modified effective average action, $\Gamma_k=\Gamma_k[\Phi_k,{\bar g},F_k]$,
is introduced through the double Legendre transform of $W_k$
\beq
\label{B15}
&&\Gamma_k[\Phi_k,{\bar g},F_k]=W_k[J,{\bar g},\Sigma]-J_i\Phi_k^i-
\Sigma_{\ell}\big({\cal L}^{(\ell)}_k(\Phi_k,{\bar g})+\hbar F_k^{\ell}\big),\\
\label{B16}
&&\Phi_k^i=\frac{\delta W_k}{\delta J_i},\quad \hbar
F_k^{\ell}=\frac{\delta W_k}{\delta \Sigma_{\ell}}- {\cal
L}^{(\ell)}_k\Big(\frac{\delta W_k}{\delta J},{\bar g}\Big),\;\;
\ell =1,2.
\eeq
From (\ref{B15}), (\ref{B16}) it follows
\beq
\label{B17} \frac{\delta\Gamma_k}{\delta\Phi_k^i}=-J_i-\Sigma_{\ell}
{\cal L}^{(\ell)}_{k,i}(\Phi_k,{\bar g}),\quad
\frac{\delta\Gamma_k}{\delta F_k^{\ell}}=-\hbar\Sigma_{\ell}.
\eeq

The modified effective average action satisfies the following functional
integro-differential equation
\beq
\label{B15a}
\!\!\!\exp\Big\{\!\frac{i}{\hbar}\Big(\Gamma_k\!
-\!\frac{\delta\Gamma_k}{\delta F_k}F_k\Big)\!\Big\}\!=\!\!
\int \!\!D\phi
\exp\Big\{\!\frac{i}{\hbar}\Big(\!S[\Phi_k\!+\!\phi,{\bar g},F_k]\!-\!
\frac{\delta\Gamma_k}{\delta\Phi_k}\phi-\!
\frac{1}{2}\frac{\delta\Gamma_k}{\delta(\hbar F_k)}{\cal L}^{(2)}_k(\Phi_k,{\bar g})
\phi\phi\Big)\!\Big\},
\eeq
where
\beq
{\cal L}^{(2)}_k(\Phi_k,{\bar g})=
\frac{\delta^2 {\cal L}_k(\Phi_k,{\bar g})}{\delta\Phi_k\delta\Phi_k}.
\eeq

In what follows below the gauge dependence of $\Gamma_k$ (\ref{B15}) is analyzed
from the point of solutions to the equation (\ref{B15a}) which
can be in principal found perturbatively in the form of loop expansions
having in this case their own specific features
due to the fact that appearing equations have
the form of Clairaut-type equations (for detailed discussions see \cite{LM}).

Let us introduce
the full sets of fields ${\cal F}_k^{\cal A}$ and sources ${\cal
J}_{\cal A}$ according to
\beq
\label{B18}
{\cal F}_k^{A}=(\Phi_k^i,\hbar F_k^{\ell})
\,,\qquad {\cal J}_{ A}=(J_i,\hbar\Sigma_{\ell}).
\eeq
From the condition of solvability of equations (\ref{B17}) with
respect to the sources \ $J$ \ and \ $\Sigma$, it follows that
\beq
\label{B19}
\frac{\de {\cal F}_k^{
C}({\cal J})}{\de {\cal J}_{B}}\,\,
\frac{\overrightarrow{\de}{\cal J}_{
A}({\cal F}_k)}{\de{\cal F}_k^{ C}} \,=\,\de^{B}_{\;A}\,.
\eeq
One can express \ ${\cal J}_{A}$ \ as a function of
the fields in the form
\beq
\label{B20}
{\cal J}_{A}
\,=\,\Big(-\frac{\de\Ga_k}{\de\Phi_k^i}\,-\,
\frac{\de\Ga_k}{\de F_k^{\ell}}\,\frac{\de L^{\ell}_k(\Phi_k,{\bar g})}{\de\Phi_k^i}
\,,\,\,-\,
\frac{\de\Ga_k}{\de F_k^{\ell}}\Big)
\eeq
and, therefore,
\beq
\label{B21}
(G^{''-1}_k)^{AC}(G^{''}_k)_{{CB}}=\de_{\;B}^{A},\qquad
\frac{\overrightarrow{\delta}{\cal J}_{B}({\cal F}_k)}{\delta{\cal F}_k^{A}}
= -(G^{''}_k)_{{AB}}\,,\qquad
\frac{\delta {\cal F}_k^{B}({\cal J})}{\delta {\cal J}_{A}}=
-(G^{''-1}_k)^{AB}\,.
\eeq
Taking into account that due to the Legendre transform
\beq
\label{B22}
\delta W_k=\delta\Gamma_k,
\eeq
the equation (\ref{B13}) in terms of modified  effective average action,
$\Gamma_k=\Gamma_k[\Phi_k,{\bar g},F_k]$,   rewrites as
\beq
\label{B23}
\delta\Gamma_k=-\Big(\frac{\delta\Gamma_k}{\delta\Phi_k^i}+
\frac{1}{\hbar}\frac{\delta\Gamma_k}{\delta  F_k}
\big({\cal L}_{k,i}(\Phi_k,{\bar g})-{\cal L}_{k,i}(\widehat{\Phi}_k,{\bar g})\big)
\Big) R^i(\widehat{\Phi}_k,{\bar g})
\delta\Psi[\widehat{\Phi}_k,{\bar g}]\;\cdot  1,
\eeq
where
\beq
\label{B24}
\widehat{\Phi}_k^i=\Phi^i_k+i\hbar (G^{''-1}_k)^{iB}
\frac{\overrightarrow{\delta}}{\delta {\cal F}_k^{B}}\;.
\eeq
From (\ref{B24}) it follows
\beq
\label{B25}
\delta\Gamma_k\Big|_{\frac{\delta\Gamma_k}{\delta {\cal F}_k}=0}=0
\eeq
the gauge independence of modified effective average action calculated on the its
extremals. This very important property of $\Gamma_k$ is found within the standard
perturbation theory accepted in Quantum Field Theory for evaluation of functional integrals.
In its turn the FRG is considered as non-perturbative approach to quantum field theories
when the  effective average action should  be found as a solution to the flow equation.
Quite recently \cite{Lav-2020} it was proved that the  effective average action
depends on gauge at every value
of the IR parameter $k$ including the fixed points making impossible physical interpretation
of any results obtained in the standard formulation of the FRG for  gauge theories.
One meets with the same drawback considering  the reformulation of the method based
on the 2PI effective action when regulators are considered as sources to composite fields
\cite{AMNS-1,Lav-alt}. In the next section we are going to introduce
an alternative flow equation and to study its $k$-dependence.

\section{Alternative flow equation and $k$-dependence}
\noindent
The flow equation in the FRG is the basic relation
describing the dependence of the effective average action on the IR
parameter $k$. Let us derive an alternative flow equation for the
2PI effective action (\ref{B15}). To do this we start with
differentiating the functional
$Z_k=Z_k[J,{\bar g},\Sigma]$ (\ref{B4})
with respect to $k$ and taking into account that only quantities
${\cal L}_{k}(\widehat{q},{\bar g})$ through the regulators $R_{k}({\bar g})$
 depend on $k$,  we obtain
the flow equation for the functional $Z_k$
\beq
\label{C1}
\pa_k Z_k=\frac{i}{\hbar}
\Sigma\;\pa_k{\cal L}_{k}(\widehat{q},{\bar g})Z_k.
\eeq
In terms of the generating functional of connected Green functions
$W_k=W_k[J,{\bar g},\Sigma]=-i\hbar\ln Z_k$, the relation
(\ref{C1}) rewrites in the form
\beq
\label{C2}
\pa_k W_k=
\Sigma\;\pa_k{\cal L}_{k}(\widehat{Q}_k,{\bar g})\cdot 1,
\eeq
In terms of the modified effective average action,
$\Gamma_k=\Gamma_k[\Phi_k,{\bar g},F_k]$
the relation
(\ref{C2}) takes the form
\beq
\label{C3}
\pa_k\Gamma_k=-\frac{\delta\Gamma_k}{\hbar\delta  F_k}
\;\pa_k{\cal L}_{k}(\widehat{\Phi}_k,{\bar g})\cdot 1,
\eeq
From (\ref{C3}) it follows
\beq
\label{C4}
\pa_k\Gamma_k\Big|_{\frac{\delta\Gamma_k}{\delta F_k}=0}=0.
\eeq
The modified flow equation
obeys the  independence on the IR parameter $k$ when it is considered on extremals,
\beq
\label{C5}
\frac{\delta\Gamma_k}{\delta F^{\ell}_k}=0, \quad \ell=1,2.
\eeq
This fact gives a hope that calculations with the modified effective average action
at the fixed points may have a physical meaning sense in contrast with the standard FRG.

In the next section we will study the gauge dependence
problem for the alternative  flow equation.

\section{Gauge dependence of alternative flow equation}
\noindent Taking into account that functions ${\cal
L}^{(1)}_{k}(h,{\bar g})$ and ${\cal L}^{(2)}_{k}(C,{\bar C},{\bar
g})$ do not depend on gauges, from (\ref{C1}) it follows that the
gauge dependence of alternative flow equation for the functional
$Z_k$ is described by the equation \beq \label{D1} \delta\big(\pa_k
Z_k\big)=\frac{i}{\hbar} \Sigma\;\pa_k{\cal L}_{k}(\widehat{q},{\bar
g})\delta Z_k, \eeq or using (\ref{B11}) as \beq \label{D2}
\delta\big(\pa_k Z_k\big)=\Big(\frac{i}{\hbar}\Big)^2
\Sigma\;\pa_k{\cal L}_{k}(\widehat{q},{\bar g})\big(J_i+ \Sigma{\cal
L}_{k,i}(\widehat{q},{\bar g}) \big)R^i(\widehat{q},{\bar g})
\delta\Psi[\widehat{q},{\bar g}]\;Z_k. \eeq We find an inspected
fact that the gauge dependence of flow equation disappears  already
when external sources to composite fields are switched off,
$\Sigma_1=\Sigma_2=0$. The alternative flow equation for the
functional $W_k$ reads \beq \label{D3} \delta\big(\pa_k W_k\big)=
\frac{i}{\hbar}\Big(\Sigma\;\pa_k{\cal L}_{k}(\widehat{Q}_k,{\bar
g})- \pa_k W_k\Big)\delta W_k, \eeq or in the form \beq \label{D4}
\delta\big(\pa_k W_k\big)= \frac{i}{\hbar}\Sigma\big(\pa_k{\cal
L}_{k}(\widehat{Q}_k,{\bar g})- \pa_k{\cal
L}_{k}(\widehat{Q}_k,{\bar g})\cdot 1 \big)\big(J_i+ \Sigma{\cal
L}_{k,i}(\widehat{Q}_k,{\bar g}) \big) R^i(\widehat{Q}_k,{\bar g})
\delta\Psi[\widehat{Q}_k,{\bar g}]\cdot  1. \eeq In terms of the
modified effective average action the gauge dependence of
alternative flow equations can be presented as \beq \label{D5}
\delta\big(\pa_k \Gamma_k\big)=
\frac{i}{\hbar}\Big(-\frac{\delta\Gamma_k}{\hbar\delta F_k}
\pa_k{\cal L}_{k}(\widehat{\Phi}_k,{\bar g})- \pa_k \Gamma_k
\Big)\delta \Gamma_k, \eeq or as \beq \nonumber &&\delta\big(\pa_k
\Gamma_k\big)=-\frac{i}{\hbar^2}\frac{\delta\Gamma_k}{\delta F_k}
\big(\pa_k{\cal L}_{k}(\widehat{\Phi}_k,{\bar g})- \pa_k{\cal
L}_{k}(\widehat{\Phi}_k,{\bar g})\cdot 1
\big)\times\\
\label{D6} &&\qquad
\times\Big(\frac{\delta\Gamma_k}{\delta\Phi_k^i}+
\frac{1}{\hbar}\frac{\delta\Gamma_k}{\delta  F_k} \big({\cal
L}_{k,i}(\Phi_k,{\bar g})-{\cal L}_{k,i}(\widehat{\Phi}_k,{\bar
g})\big) \Big) R^i(\widehat{\Phi}_k,{\bar g})
\delta\Psi[\widehat{\Phi}_k,{\bar g}]\;\cdot  1.
\eeq
From
(\ref{D5}), (\ref{D6}) it follows
\beq
\label{D7}
\delta\big(\pa_k \Gamma_k\big)\Big|_{\frac{\delta\Gamma_k}{\delta
F_k}=0}=0.
\eeq
Therefore, the flow equation is gauge independent on extremals,
\beq
\label{D8}
\frac{\delta\Gamma_k}{\delta F^{\ell}_k}=0,
\eeq
and as well as additionally obeys the  independence on the IR parameter $k$.
These facts give a possibility for  consistent application of the proposed
quantization procedure in gauge theories to obtain physical meaning results.
It is interesting to note that the gauge independence of $\Gamma_k$
found as a solution
to the flow equation is expected with use only a part of
the equations of motion (\ref{D8}). In its turn
the  independence of the flow equation on the IR parameter $k$
at the extremals (\ref{D8}) can be considered as a signal about
$k$-independence of $\Gamma_k$ at the fixed points.

\section{Summary}
\noindent
We have studied basic properties of the flow equation in gravity theories
within reformulation of
the standard FRG  \cite{Wet1,Wet2}
when following ideas of \cite{LSh} the 2PI effective action
with composite fields being densities of the regulator action is
introduced. In contrast with the standard FRG \cite{Wet1,Wet2} and
alternative approaches \cite{AMNS-1,Lav-alt} the proposed reformulation
leads to the 2PI effective average action which possesses standard properties
of gauge dependence  in the perturbation theory
and satisfies the alternative flow equation being   $k$-
and gauge independent at  extremals. Speaking about possible reformulations of the
standard FRG it is necessary to notice the papers \cite{Morris1,Morris2}
where the quantization procedure is based on gauge invariant regularization
of an initial classical action which guarantees the BRST symmetry and the gauge independence
of $S$-matrix elements. Unfortunately absence of explicit procedure to arrive the gauge
invariance of regularized initial action does not allow to consider this approach
as a consistent quantization method \cite{Lav-RG-BV}.
Therefore at the moment we state that the proposed reformulation of the standard
FRG should be considered
as non-perturbative quantization of gauge theories successfully passing
through the gauge dependence tests.

\section*{Acknowledgments}
\noindent
The work  is supported by the Ministry of  Education of the Russian Federation,
project FEWF-2020-0003.

\end{document}